\documentclass[superscriptaddress,twocolumn]{revtex4}
\usepackage{amsmath,lscape,epsfig}
\usepackage{graphicx}

\def\beq{\begin{equation}}
\def\eeq{\end{equation}}
\def\beqa{\begin{eqnarray}}
\def\eeqa{\end{eqnarray}}
\def\ban{\begin{eqnarray*}}
\def\ean{\end{eqnarray*}}
\def\bi{\begin{itemize}}
\def\ei{\end{itemize}}

\begin{document}

\title{Phase transition of the nucleon-antinucleon plasma at different 
ratios }

\author{ A. Delfino, M. Jansen }
\affiliation {Depto. de F\'{\i}sica - Universidade Federal
Fluminense - CEP 24210-346 - Niter\'oi RJ, Brazil}
\author{V. S. Tim\'oteo}
\affiliation{Centro Superior de Educa\c c\~ao Tecnol\'ogica, 
 Universidade Estadual de Campinas, 
13484-332 Limeira, SP, Brazil}

\begin{abstract}
We investigate phase transitions for the Walecka 
model at very high temperatures. 
As is well known, depending on the parametrization of this model and 
for the particular case of a zero chemical potential ($\,\mu\,$), 
a first order phase transition is possible \cite{theis}. We investigate 
 this model for the case in which $\,\mu \ne 0\,$. 
It turns out that, in this situation, phases with different values of 
antinucleon-nucleon ratios and net baryon densities may coexist. We present 
the temperature versus antinucleon-nucleon ratio as well as the temperature 
versus the net baryon density for the coexistence region.  The 
 temperature versus chemical potential phase diagram is also presented. 

\end{abstract}

\maketitle

\vspace{0.0cm} PACS number(s):
{21.65.+f,24.10.Jv,21.30.-x,21.60.-n}
\vspace{0.50cm}


Finite nuclei and nuclear matter properties have been reasonably well 
described by the Walecka model \cite{walecka}. 
This renormalizable model employs nucleons and mesons ($\sigma$ and $\omega$) 
as the degrees of freedom. The sources for the fields are the scalar 
($\rho_{s}$) and vector ($\rho$) densities associated with the Lorentz scalar 
($S$) and vector ($V$) interactions.   
It is well known that this model, after the fitting of the experimental 
infinite nuclear matter binding energy at the saturation density, gives 
for this system a high bulk incompressibility, as well as a high spin-orbit
splitting energy, when applied for finite nuclei. In order to overcome this 
problem, nonlinear relativistic hadronic models which include cubic and 
quartic scalar mesonic sef-couplings have been proposed \cite{boguta}. The 
preliminary success of this kind of nonlinear Walecka model, estimulated the 
proposal of different parametrizations to better describe nuclear matter 
properties \cite{nl}. Still aiming at refining its predictions for nuclear 
matter and neutron stars properties, new kinds of  models including 
scalar(vector) self-coupling interactions added to usual nonlinear Walecka 
models \cite{boguta,nl} have been constructed \cite{ohio1}. Effective hadronic 
models with density dependent coupling constants have also been proposed
\cite{density}. 

In general, (non)linear Walecka models models 
\cite{walecka,boguta,nl,ohio1,density} 
have been investigated under extreme density and temperature regimes.  
Most of such studies focus on the hadronic quark-gluon plasma phase 
transition, which is not the aim of our present work. Here, our purpose is 
to extend the very interesting work of Theis et al \cite{theis}. 
In their study \cite{theis}, the Walecka model was investigated in the extreme 
high temperature regime in which the chemical 
potential ( $\,\mu\,$ )is zero. This leads to a situation in which the net 
baryon density ($\,\rho\,$) also vanishes. In this case, the number of 
nucleon and anti-nucleon are the same. This study showed that, in 
this nucleon-antinucleon plasma at very high temperature, a phase 
transition exists. 

In this work, we will investigate the Walecka model under an extreme 
temperature regime, but allowing the number of anti-particles and particles to 
be different in the two phases it can lead to. 
The question we pose here is whether by allowing different ratios of 
nucleons-anti-nucleons a phase transition scenario is still present, as in the 
particular case $\rho=0$, still in the so called 'no-sea' approximation. That 
is, we consider explicitly only the valence Fermi and Dirac sea states. 
Let us here remark that such an approximation, as investigated carefully by 
the Ohio group \cite{ohio2}, becomes good at low energies, where the 
contributions of the Dirac sea can be renormalized in effective coupling 
constants. Of course, a rigorous way to perform our investigation would be to 
include vacuum polarization effects in the Walecka model for high excitation 
energies, where the anti-particles play an essential role. This treatment, however, 
is beyond the scope of this work and we perform an exploratory investigation even though 
the validity of the 'no-sea' approximation in this regime may not be granted. 

Nowadays, high energy experiments reveal evidence of nuclear systems with very 
small baryonic density in the study of particle yields measured in central 
$Au-Au$ collision at RHIC. Different experimental analysis (STAR, PHENIX, 
PHOBOS, BRAHMS) furnish the antiproton-proton ratio 
$\bar p /p \approx 0.65$ for a temperature of $174MeV$ and a chemical baryonic 
potential of $46MeV$ estimated from thermal models to fit 
antiparticle-particle ratios \cite{brauna}. High energy $Pb-Pb$ collisions 
show that this ratio reach values around 0.9 \cite{braunb}. If relativistic 
hadronic models are to be used in the description of the multiplicity observed 
today in ultra-relativistic heavy ion collisions \cite{detlef,debora}, the 
behavior of these models at high temperature regimes may be of importance. 

The thermodynamics of the Walecka model may be given, for finite temperature, 
in terms of the energy density and pressure functionals, 
\begin{eqnarray}
 {\cal E}\, &=&\, \frac{C_{\omega}^{2}}{2M^{2}}\,\,\rho_{b}^2 + 
 \frac{2C_{\sigma}^{2}}{2M^{2}}\,\,\rho_{s}^2 \nonumber \\
 &+&\frac{\gamma}{2\pi^{2}}\,\,\int k^{2} dk E^{\ast}(k)(n_{k}+\bar{n}_{k}) 
 \label{energy}
\end{eqnarray}
and
\begin{eqnarray}
  p\,&=&\, 
\frac{C_{\omega}^{2}}{2M^{2}}\,\,\rho_ {b}^2 - 
 \frac{2C_{\sigma}^{2}}{2M^{2}}\,\,\rho_{s}^2 \nonumber \\
 &+&\frac{\gamma}{6\pi^{2}}\,\,\int \frac{k^{4} dk}{E^{\ast}(k)}
(n_{k}+\bar{n}_{k}) , 
\label{pressure}
\end{eqnarray}
where 
\begin{eqnarray}
 \rho_{b} = \rho - \bar \rho  ,
\label{baryondensity} 
\end{eqnarray}
\begin{eqnarray} 
\rho =  \frac{\gamma}{ 2 \pi^{2}}\int k^{2} dk \; n_{k}  \;\; ,
\label{particledensity} 
\end{eqnarray}
\begin{eqnarray} 
\bar \rho = 
  \frac{\gamma}{ 2 \pi^{2}}\int k^{2} dk \; \bar n_{k} \;\; ,
\label{antiparticledensity}  
\end{eqnarray}
\begin{eqnarray}
\,\,\rho_{s}\,=\,M^{\ast}\frac{\gamma}{2 \pi^2}
\int \frac{k^{2}dk}{ E^{\ast}(k)} \; (n_{k}\,+\,\bar 
n_{k}) \;\; .  \label{scalardensity}   
\end{eqnarray}
and  
In the  expressions above, $\, \gamma\,$ is the degeneracy factor 
( $\,\gamma = 4\, $ for nuclear matter and $\, \gamma = 2\, $ for neutron 
matter ), $n_{k}$ and $\bar n_{k}$ stand for the Fermi-Dirac distribution for 
baryons and antibaryons respectively, with arguments 
$ (E^{\ast} {\mp} \nu)/T $ . $E^{\ast}(k)$ is given by
$\,   E^{\ast}(k)\, =\, ( k^2 + M^{\ast 2} )^{1/2}\,$, 
whereas an effective chemical potential, which preserves the number of baryons 
and antibaryons in the ensemble, is defined by 
$\,\,\nu\,=\,\mu\, -\, V $ , $\,V=C_{\omega}^{2} \rho_{b} /M^{2} $, 
where $\mu$ is the thermodynamic chemical potential.
The solution for the equation of state is obtained explicitly through the 
minimization of $\,{\cal E }\,$ relative to the scalar field, or equivalently 
to $\,m^{*}=M^{*}/M$, 
\begin{eqnarray}
 f(\frac{M^{*}}{M})\,\,\,=1\,\,-\,\,\frac{M^{*}}{M}\,\,-\,\,\frac{C_{s}^2}{M^3}
\,\,\rho_{s}  \,\,=\,\,0\,\,. \label{gap}
\end{eqnarray} 
 
This equation, known as the gap equation, has to be solved self-consistently 
with Eqs. \ref{energy}-\ref{pressure} and provides the
basis for obtaining all thermodynamic quantities in
the mean field approach we are using.

Usually, the constants $C_{\sigma}^{2}$ and $C_{\omega}^{2}$ are determined 
in favor of the experimental nuclear matter binding energy ($16MeV$) at 
the saturation density ($\rho_{b}=\rho_{o}=0.15fm^{-3}$), at $T=0$. At 
finite temperature, this model predicts a critical liquid-gas behavior as 
a van der Waals EoS. Its critical temperature is around $18MeV$ 
\cite{walecka}. For this temperature, the antinucleon-nucleon ratio, 
given by $R= \, \bar \rho /\rho$ is negligible. Only when the temperature 
gets higher, the antinucleon density starts to take significant values. 

\begin{figure}[h]
\centering
\includegraphics[height=6cm, width=7cm]{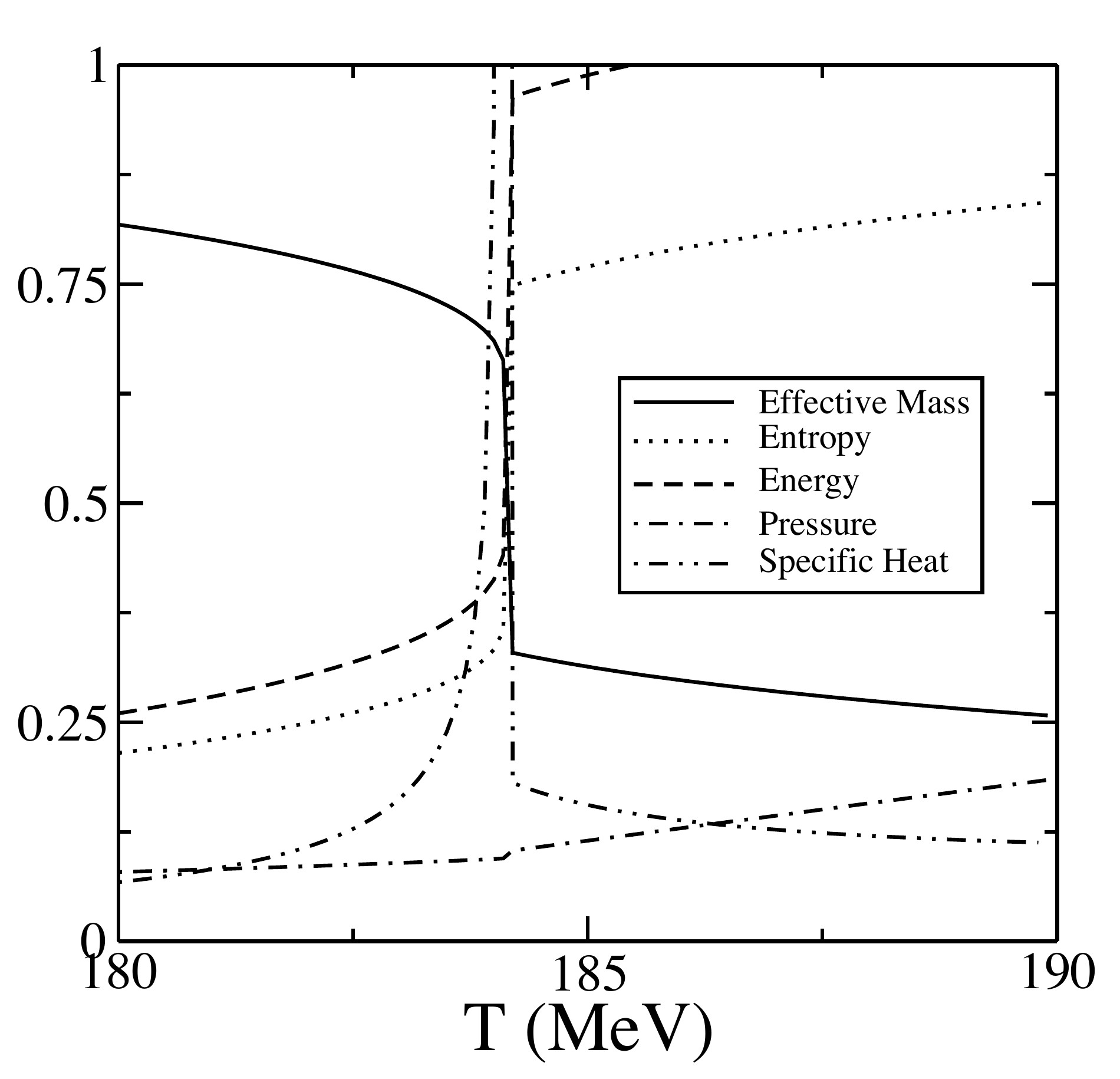}
\caption{The effective mass, entropy, energy density, pressure, and the 
specific heat for $R=1$ as a function of the temperature. The mass is in units 
of the nucleon mass and the entropy, energy, and pressure are in 
Stefan-Boltzmann units.}
\label{fig1}
\end{figure}

In a very interesting work \cite{theis}, the Walecka model was studied 
in the extreme situation in which $\,R\,=\,1\,$. In this case, 
$\rho_{b}=0$ and $\nu \,=\,\mu \, = \,0$. This study showed that, in 
this nucleon-antinucleon plasma at very high temperature, a phase 
transition exists. It is also remarkable to note that the order of the 
phase transition itself becomes dependent on the $C_{\sigma}^2$ versus 
$C_{\omega}^2$ space parameter. By small changes on these parameters (which 
means to change by a few percent the nuclear matter binding energy and 
the saturation density), the 
phase transition changes from first to second order. If we take 
the values of  $C_{\sigma}^2 = 359.35 $ and  $C_{\omega}^2 = 275.12 $, 
fitting the infinite nuclear matter binding energy as $16MeV$ at a density of 
$0.15fm^{-3}$, the phase transition is of first order. 
The findings of Ref. \cite{theis} may be summarized in Fig. \ref{fig1}. 

In Fig. \ref{fig1} we see that the nucleon effective mass decreases 
abruptly at 
$T \approx 184 MeV$. Since all other thermodynamic quantities are 
dependent on $m^{*}$, this effect also manifests itself in energy density, 
pressure, and specific heat, calculated as the temperature first derivative 
of the energy density. The fast increase of the entropy is a clear signal 
of a first order phase transition. Since the baryonic density is kept 
constant $\rho_{b}=0$,  the order parameter should be the entropy. 
It is a very curious constrained system, interpreted as a 
nucleon-antinucleon plasma \cite{theis}.    

Here, following \cite{theis}, we start to study the Walecka model 
in different constrained cases. The main question is to understand 
the behavior of  some thermodynamical quantities when  
the ratio $\,R\,=\, \bar \rho / \rho \,$ varies.  Now, all quantities 
will have contribution from the baryonic density, contrary to Fig. 
\ref{fig1} where $\rho_{b}=0$. 

\begin{figure}[h]
\centering
\includegraphics[height=6cm, width=7cm]{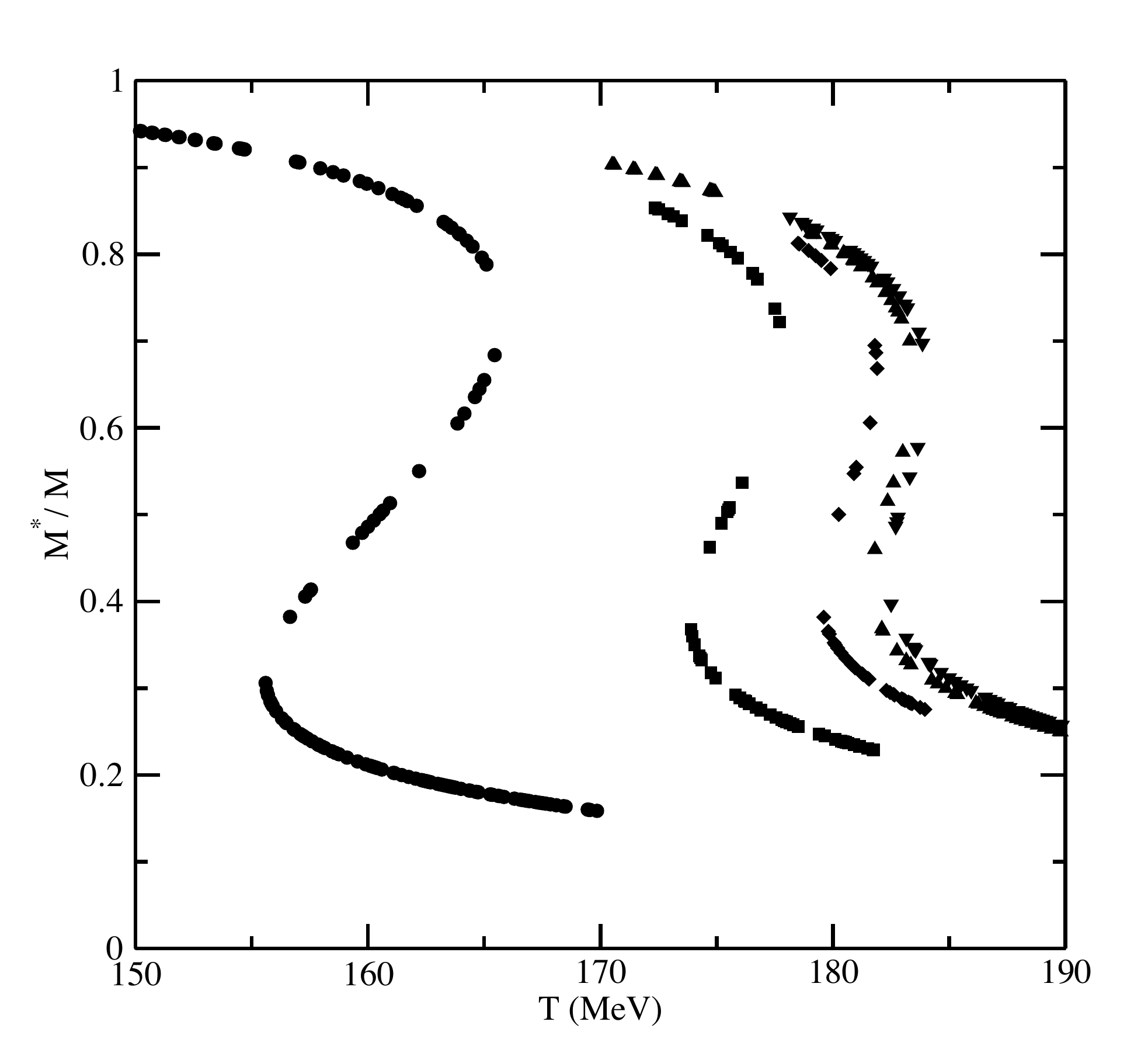}
\caption{The effective mass as a function of the temperature for several values of $R$. From the left to the right: $R=0.1, 0.3, 0.5, 0.7, 0.9$.}
\label{fig2}
\end{figure}

Let us now remark that we have fixed numerically the ratio within a 
precision of one part in a thousand. It means that the set of equations (1-7) is 
solved self-consistently with $ R -R/1000 \le \,\,R \le\,\,  R + R/1000$. 
In figure \ref{fig2} we present  $\,M^{*}/M\,$ as a function of the temperature 
for different values of $R$. 

\begin{figure}[h]
\centering
\includegraphics[height=6cm, width=7cm]{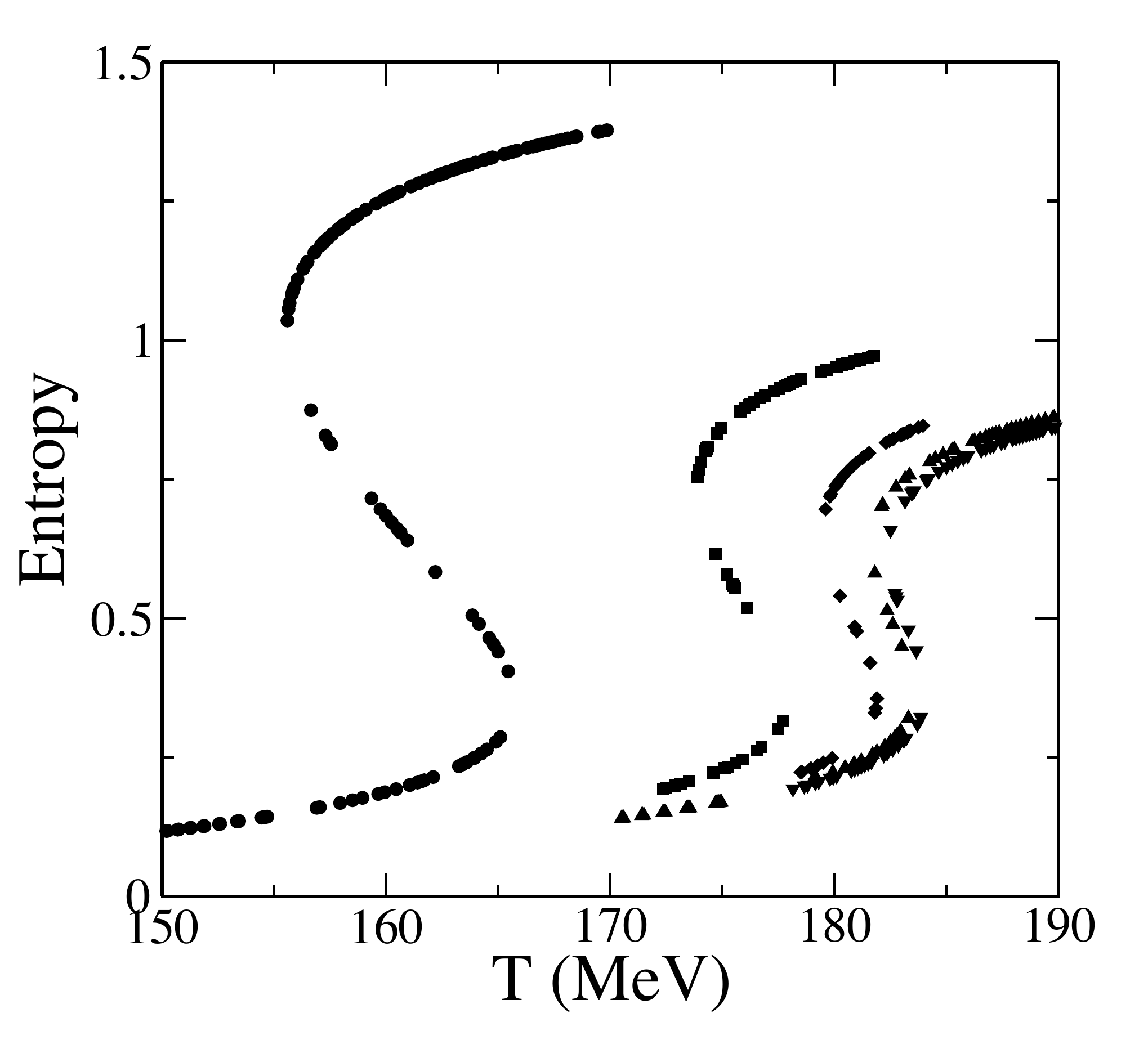}
\caption{The entropy, in Stefan-Boltzmann units, as a function of the
 temperature for several values of $R$. The values of $R$ are the same of
 figure \ref{fig2}.}
\label{fig3}
\end{figure}

For the same values of $R$ investigated in Fig. \ref{fig2}, 
the entropy 
behavior as a function of the temperature is shown in Fig. \ref{fig3}  

\begin{figure}[h]
\centering
\includegraphics[height=6cm, width=7cm]{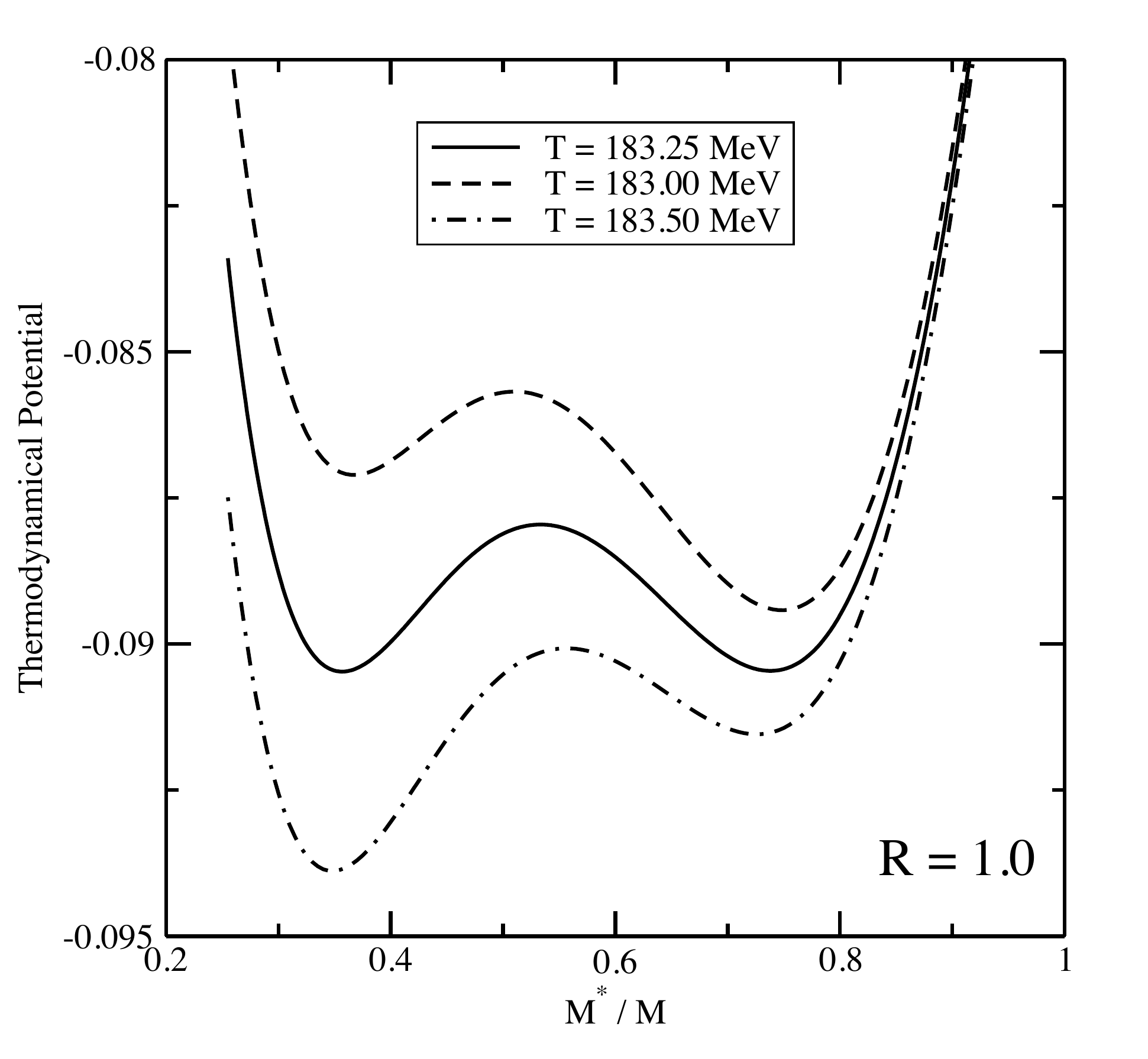}
\caption{Thermodynamical potential, in units of the Stefan-Boltzmann pressure, 
as a function of the nucleon effective mass, for temperatures around the critical 
temperature in the case where $\mu = 0$ ($R=1$).}
\label{fig4}
\end{figure}

Both the nucleon effective mass and the entropy follow the same abrupt 
decreasing (increasing) behavior one sees for the 
particular case $R=1$.  
From these figures we see that, as the ratio $\bar \rho / \rho$ decreases,
the temperature in which an abrupt behavior arises also decreases, 
but keeping the same character.  
In principle, there is an important 
difference between the system described by Fig. \ref{fig1} and those of 
Figs. \ref{fig2} and \ref{fig3}. In the first, 
$R=1$ and $\rho_{b}=0$ along the temperature variation. In the second, 
the ratios are kept constant while $\rho_{b}$ varies. 
Visually, Figs. 2-3 suggest phase coexistence at some temperature range. 
According to the Gibbs criteria, if one has phases $1$ and $2$, a phase coexistence
arises when $p_{1} \,=\,p_{2}$, $\mu_{1} \,=\,\mu_{2}$ and 
$T_{1} \,=\,T_{2}$. The critical temperature is achieved when, 
above that, no more phase coexistence is possible. For the case $R=1$, a 
phase coexistence exists at $T=183.25MeV$ and $\,\mu=0\,$. 
It means that, if the temperature increases or decreases from 
this value, the phase coexistence disappears as we can see in Fig. \ref{fig4}, 
where the minima of the thermodynamical potential have different values for the same temperature.

This kind of investigation was performed, for example, by
Asakawa and Yazaki \cite{yazaki} when working with the NJL model at finite temperature, 
where a phase transition is also present. Following their procedure, we can verify whether 
the system exhibits phase coexistence or not.

Surprisingly, despite the signals of phase coexistence shown by 
Figs. \ref{fig2}-\ref{fig3}, we did not find 
phase coexistence for a single value of $R$ when $\,R\,\ne\,1\,$. 
In this case, when the system was in thermal ($T_{1}\, =\, T_{2}$) 
and mechanical  ($p_{1}\,=\,p_{2}$) equilibrium, chemical equilibrium 
($\mu_{1}\,=\,\mu_{2}$) was not found. On the other hand, along the 
apparent phase coexistence region signalized in Figs. 
\ref{fig2}-\ref{fig3} by three 
different values of $M^{*}/M$ and entropy for the same temperature, we
found a stable phase (global minimum of the thermodynamical potential). 
It indicates the presence of stable phases up to the start of the backbending 
of the curves, and new stable phases by any increase of the temperature, but
with a dramatic decrease (increase) in the effective mass (entropy). 
Such a phase transition without phase coexistence is rare in physics. 
Therefore, we decided to leave the fixed ratio scenario to analyse what happens 
if the system becomes free of such constraint.  

\begin{figure}[h]
\centering
\includegraphics[height=6cm, width=7cm]{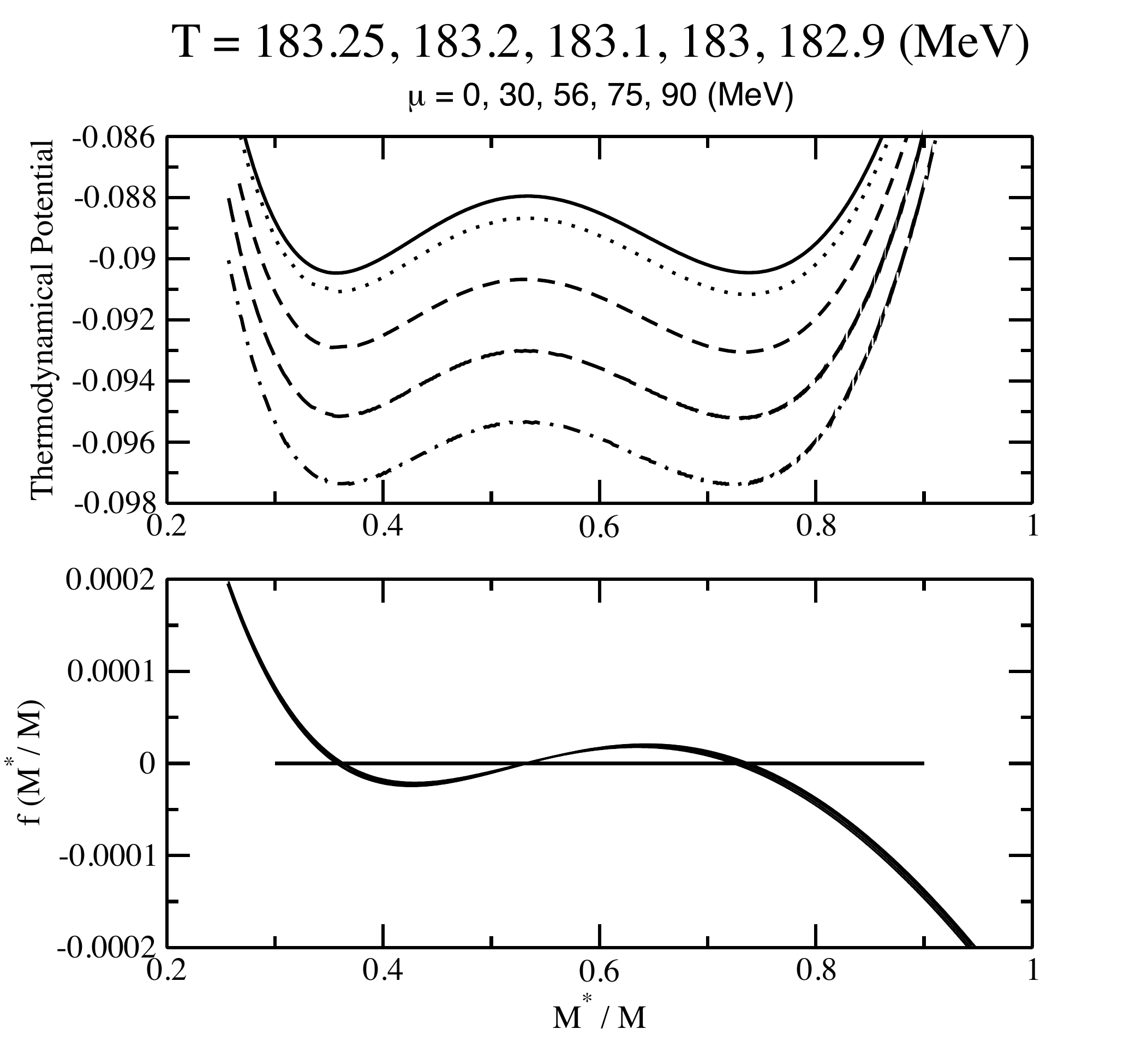}
\caption{The same as in the previous figure, but showing the phase coexistence for 
temperatures between $T=183.25 MeV$ (upper curve in top panel)
and $T=182.90 MeV$ (lower curve in botton panel). The values of $\mu$
are those that make the two minima to be have the same value for each temperature. 
Note that, now, $R$ is not constrained and have a different value for each 
temperature and chemical potential.} 
\label{fig5}
\end{figure}

Next, we proceed to investigate the phase coexistence 
in the Walecka model still at high temperature but without any fixed 
$\,R\,$ constraint. In Fig. \ref{fig5} we display the thermodynamical 
potential density $\,-p\,$ and  f($M^{*}/M$)  given by Eq. (\ref{pressure}) 
and Eq. (\ref{gap}) respectively. 

\begin{figure}[h]
\centering
\includegraphics[height=6cm, width=7cm]{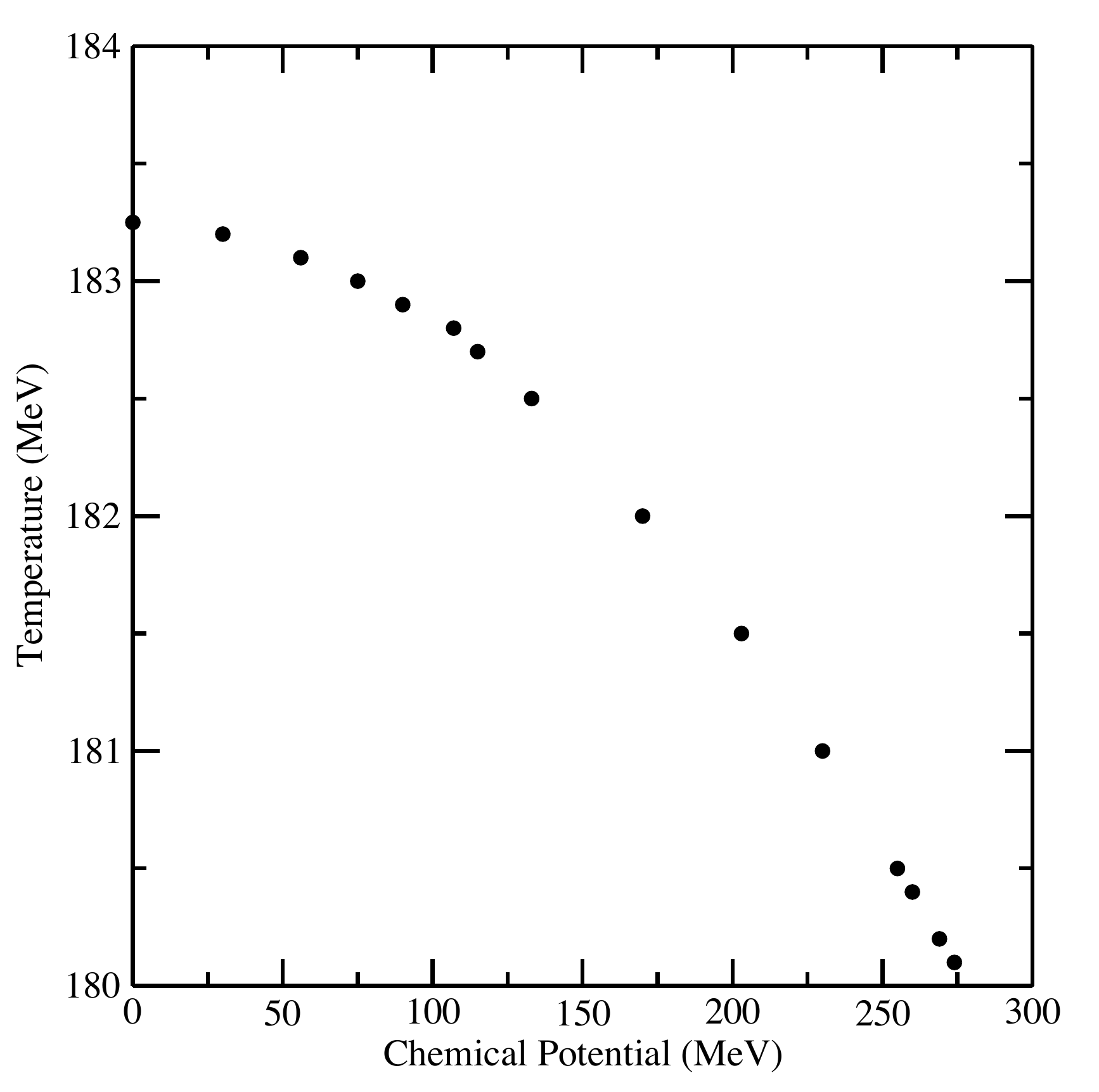}
\caption{Temperature versus baryon chemical potential for the coexistence 
phase region.}
\label{fig6}
\end{figure}

In the curves presented, the Gibbs 
criteria $T_{1}\, =\, T_{2}$, $p_{1}\,=\,p_{2}$) and $\mu_{1}\,=\,\mu_{2}$ 
are clearly satisfied. As we continue to decrease the temperature, the Gibbs 
criteria can be fulfilled since the chemical potential $\,\mu\,$ increases.  
The curves for the thermodynamical potential for 
$\,T\,<\,182.9MeV\,$ (not shown) becomes flatter as the temperature 
decreases. This happens until $\,T\,\approx 180MeV\,$ 
and $\mu \approx 274MeV\,$.   
Therefore, the coexistence region for the Walecka model, with the coupling constants 
given previously and without any $R$ constraint, 
is $180.1\,\le\,T\,\le\,183.25MeV\,$ with 
$0\,\le\,\mu \,\le\,274MeV\,$.  
The phase diagram $\,T\,\times\,\mu$ is given in Fig. \ref{fig6}. 

\begin{figure}[h]
\centering
\includegraphics[height=6cm, width=7cm]{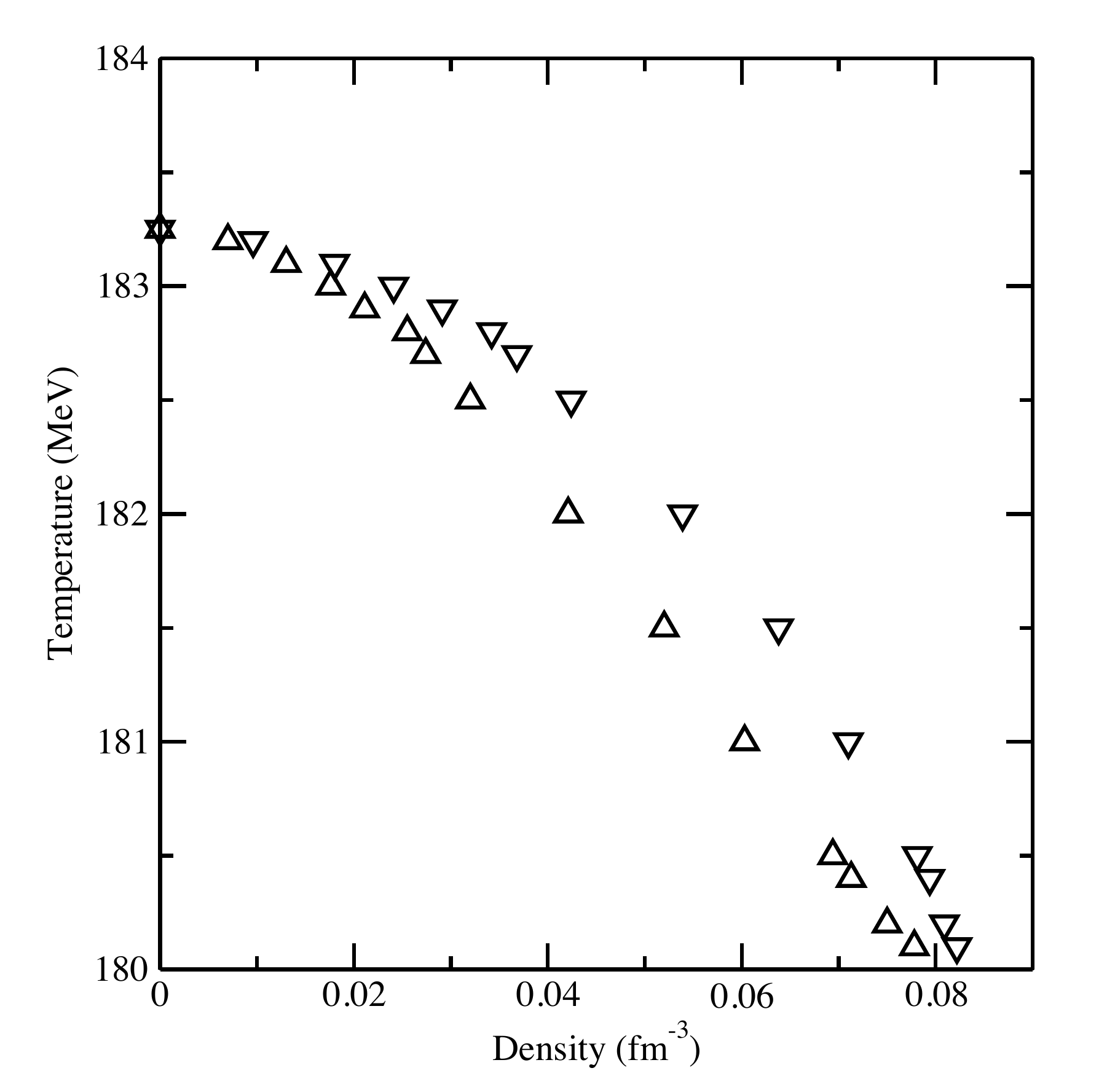}
\caption{Temperature versus net baryon density  for the phase coexistence region.}
\label{fig7}
\end{figure}

Now, we have the values of $M^{*}/M$ which allow the phase coexistence 
for the phase diagram of Fig. \ref{fig6}. With these, we can extract the 
net baryon 
densities $\,\rho\,$ as a function of $\,T\,$ which allows phase 
coexistence. The results are presented in Fig. \ref{fig7}.  

\begin{figure}[h]
\centering
\includegraphics[height=6cm, width=7cm]{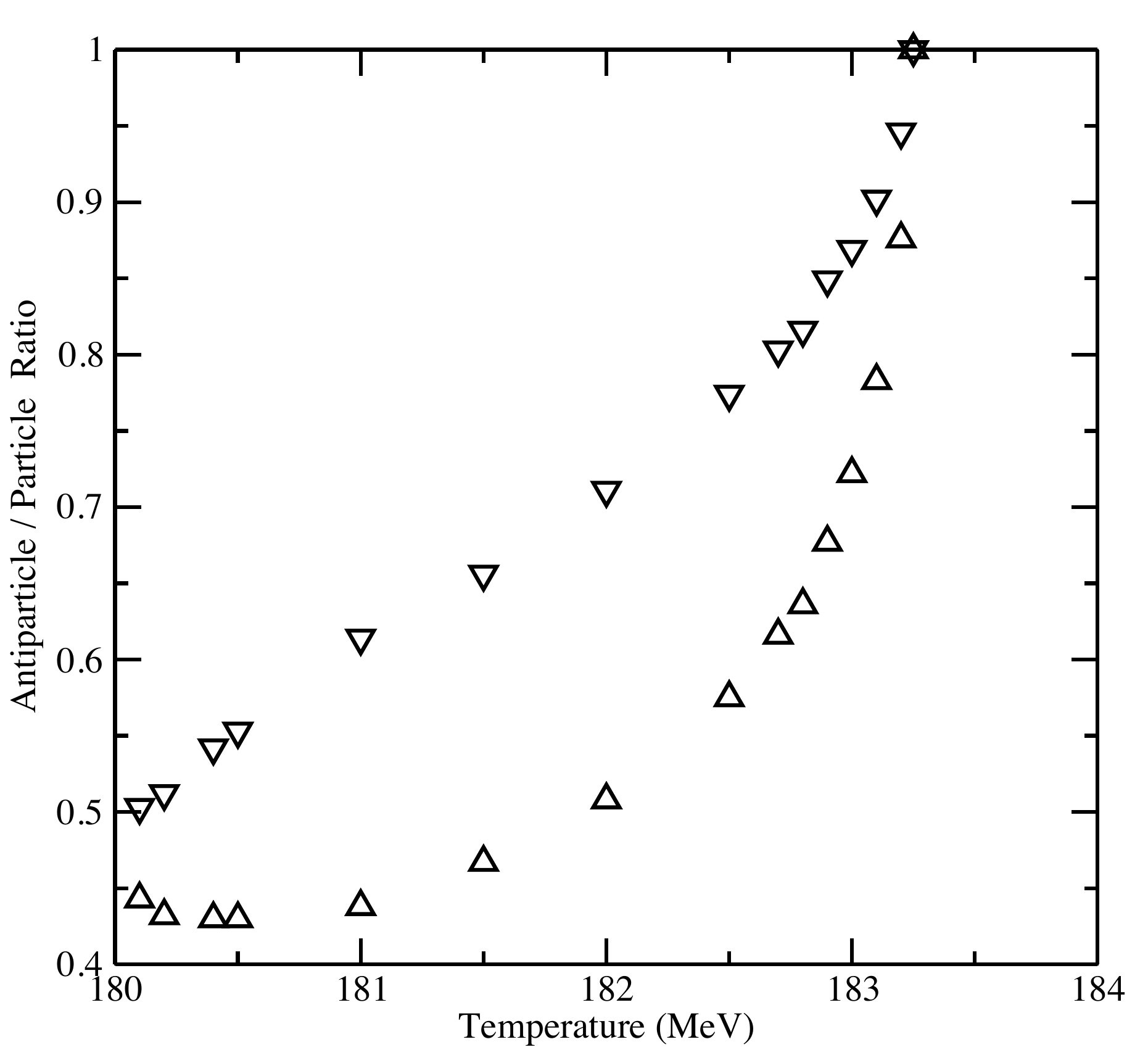}
\caption{Antinucleon-nucleon ratio ($R$) versus temperature 
for the phase coexistence region.}
\label{fig8}
\end{figure}

Following the same procedure, in Fig. \ref{fig8} we show  the different 
ratios $\,R\,$ for which the system affords coexistence. It is interesting to 
observe that no coexistence exists if one of the ratio is not greater than 
$\,1/2$. 
As we can see, only in the 
particular case $\,R=\,1\,$, which corresponds to $\,\rho_{b}\,=0\,\mu$, 
there is a phase coexistence allowing only one fixed ratio. 

In short, we start by studying  the Walecka 
model at high temperature regime constraining  the 
anti-nucleon-nucleon ratio ($R$) to be constant but not equal to one as has 
been done by Theis et al. \cite{theis}. We have seen that the visual signals 
of phase transition for the effective nucleon mass ($M^{*}$) and entropy 
($S$) versus temperature ($T$) for the cases $R \ne 1$ are tipically the same 
of those obtained in the case $R=1$. By this we mean an abrupt 
decrease (increase)  of both ($M^{*}$ and $S$ ) for $T > 180MeV$. 
Surprisingly, however, we could 
not find for $R \ne 1$, contrary to the $R=1$ case, a phase coexistence 
signature by using the Gibbs criteria. Phase transition without phase 
coexistence is rare to occur in physics. As far as we know, they are 
theoretically predicted for the anomalous two-dimension Kosterlitz-Thouless 
phase transition \cite{thouless}, as an example. 
Since we could not find a clear signature for phase coexistence at fixed 
$\,R \ne 1\,$ we 
have investigated phase coexistence for non-constrained values of $R$. Our 
study shows that the only phase coexistence regime for a fixed $R$ occurs at 
$R=1$ or, what is the same, $\mu =0$. In the the Walecka model parametrization 
we have used, coexistence phases occur approximately in the interval 
$\,180 \le \,T \, \le 183.25 MeV$ and different values of $R$ at 
each phase are needed.  

If one uses hadronic models \cite{debora,detlef} to study high energy 
heavy-ion collisions, the models face low density and high temperature 
regime. In the chemical freeze-out, anti-nucleon-nucleon are
produced at ratios which depends on the center-of-mass energy \cite{cleymans}. 
Although the Walecka model is to simple to deal with, it may anticipate roughly
what can happen with more realistic hadronic models. Theis et al 
\cite{theis} analysed ($\mu=0$) that, depending on its parametrization, 
the Walecka model shows first or second order phase transition. The same 
happens with different hadronic models \cite{delfino}. Therefore, it is to 
be expected that models  (with $\mu=0$) which present first order phase 
transition would follow a behavior similar to the Walecka model 
(Figs. \ref{fig6} -\ref{fig8}) regarding phase coexistence regions. 
Therefore, depending on 
 the values of $T$ and $ \mu$ used to fit the freeze-out data, 
the chosen hadronic model may be dealing with hadronic phase coexistence 
region, as depicted in figures \ref{fig6}-\ref{fig8}. 

\section*{ACKNOWLEDGMENTS}

This work was partially supported by CNPq and CAPES (Brasil).
V.S.T. would like to thank FAPESP and FAEPEX/UNICAMP for 
financial support.

\end{document}